\documentclass{article}
\usepackage{spconf}
\usepackage{amsmath,graphicx}

\usepackage{enumitem}
\usepackage{amsfonts}
\usepackage{booktabs}
\usepackage[caption=false]{subfig}
\usepackage{url}
\DeclareMathOperator{\softmax}{{\mathsf{softmax}}}
\usepackage{color}
\usepackage{multirow}

\title{Learning to Count Words in Fluent Speech\\ enables Online Speech Recognition}
\name{George Sterpu, Christian Saam, Naomi Harte}
\address{Sigmedia Lab, ADAPT Centre, Trinity College Dublin, Ireland}

\begin{document}
%
\maketitle
\begin{abstract}
Sequence to Sequence models, in particular the Transformer, achieve state of the art results in Automatic Speech Recognition. Practical usage is however limited to cases where full utterance latency is acceptable. In this work we introduce Taris, a Transformer-based online speech recognition system aided by an auxiliary task of incremental word counting. We use the cumulative word sum to dynamically segment speech and enable its eager decoding into words. Experiments performed on the LRS2, LibriSpeech, and Aishell-1 datasets of English and Mandarin speech show that the online system performs comparable with the offline one when having a dynamic algorithmic delay of 5 segments. Furthermore, we show that the estimated segment length distribution resembles the word length distribution obtained with forced alignment, although our system does not require an exact segment-to-word equivalence. Taris introduces a negligible overhead compared to a standard Transformer, while the local relationship modelling between inputs and outputs grants invariance to sequence length by design.
\end{abstract}
\begin{keywords}
Online ASR, word segmentation
\end{keywords}

\section{Introduction}

Having a natural conversation with a computer has fascinated humankind for a long time. A key ingredient of this ambition is granting computers the ability to recognise spoken words with minimum latency. This allows a more interactive communication, where the computer is able to interrupt a speaker to acknowledge or ask for clarifications.

Despite the remarkable progress in end-to-end automatic speech recognition technology based on sequence to sequence neural network architectures~\cite{chiu2018}, an unresolved issue is reducing the latency from full utterances down to a few words. This sentence-level, or offline conditioning, is a fundamental barrier to online decoding. 

Humans develop the ability to segment words in continuous speech from the earliest stages of life~\cite{jusczyk1995}. There is evidence that we integrate a set of acoustic, phonetic, prosodic, and statistical cues in order to segment words in fluent speech~\cite{johnson2001}. This leads us to ask whether the ability to segment speech into \emph{word} units with a neural network offers the potential to help crack the challenge of decoding online. This approach would take advantage of the monotonicity of speech, allow the network focus on local properties, and remove the offline conditioning.

To this end, we introduce Taris, a Transformer-based system for online speech recognition that learns to model the local relationships between text and audio in speech, relaxing the global conditioning constraint of the original model. We achieve this through self-supervision by introducing an auxiliary word counting task which facilitates the segmentation of speech. Taris allows efficient minibatch training and introduces a negligible overhead compared to the original Transformer model, without trading off the recognition accuracy. We make our software implementation publicly available\footnote{https://github.com/georgesterpu/Taris}.

\section{Background}
\label{sec:background}

A major technical challenge in online speech decoding is formulating the problem in a fully differentiable framework. Previous attempts include the Recurrent Neural Network Transducer~\cite{graves2012,graves2013,rao2017, battengerg2017}, Neural Transducer~\cite{jaitly2016, sainath2018}, segmental conditional random fields~\cite{Beck2018, tang2017}, hard monotonic attention~\cite{luo2017,raffel2017}, segment attention~\cite{mocha, Fan2019, Hou2020}, or triggered attention~\cite{moritz2019}. However the models made use of dynamic programming, training in expectation, or policy gradients, and the authors report training difficulties. Our work retains the segment attention design, but tackles the problem of speech segmentation from a different angle. By learning to count words through self-supervision, we introduce a mechanism that allows end-to-end training using only backpropagation.

Recent proposals in online speech recognition address this challenge by assuming one sub-word unit per segment~\cite{li2019acs, dong2019cif}, or discover an inventory of sub-word units~\cite{drexler2020}, a concept previously explored in machine translation~\cite{Kreutzer2018}. Our focus in this work is on word units. In English, words allow a monotonic and bijective mapping between their acoustic and symbolic representations, however these properties do not hold at the sub-word level due to the highly complex spelling rules in English orthography. Moreover, words can be counted in a deterministic way, which allows us to introduce a self-supervision word counting task without requiring new labels.

The sequence to sequence (seq2seq) architecture was proposed in \cite{attention_seq2seq, sutskever2014}. An Encoder transforms a variable length input sequence into a sequence of latent representations, and a Decoder maps the latent sequence onto a target sequence of a different length, aiming to establish a soft-alignment between elements of the inputs and the targets. In attention-based seq2seq networks, the conditional dependency of each output token on the entire input sequence prohibits online decoding. Yet, it has been shown that, once convergence is reached, there are predominantly local relationships between the output tokens and the audio representations in speech~\cite{chorowski_neurips2015, chan_icassp2016}. Therefore, potentially incurring no loss in accuracy, a local conditioning of the outputs on the inputs would break the offline limitation and reduce the algorithmic latency. The new goal is to learn robust associations between input and output subsequences which stand for the same linguistic concepts.

\begin{figure}[t]
        \centering
        \subfloat[Recurrent Neural Network]{
           \includegraphics[width=0.40\linewidth]{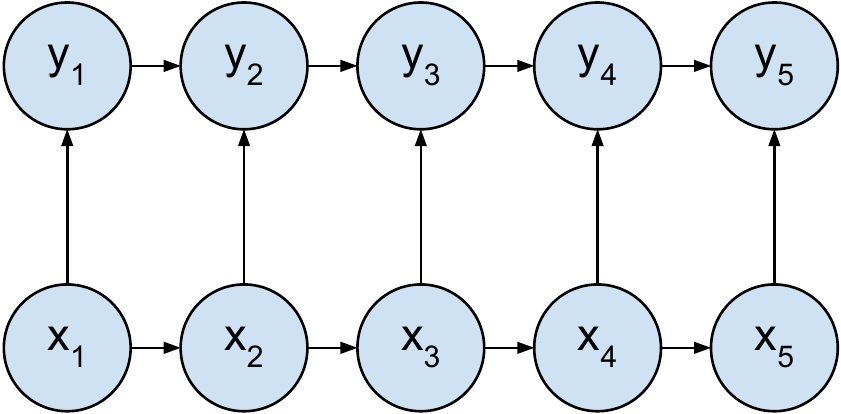}
           \label{fig:con:rnn}
        }
        \hspace{1cm}
        \subfloat[Offline Transformer]{
        \includegraphics[width=0.40\linewidth]{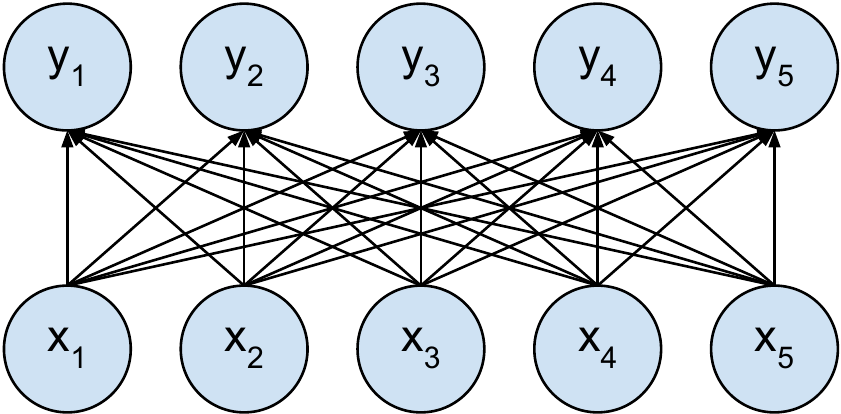}
           \label{fig:con:off}
        }

        \subfloat[Causal Transformer]{
           \includegraphics[width=0.40\linewidth]{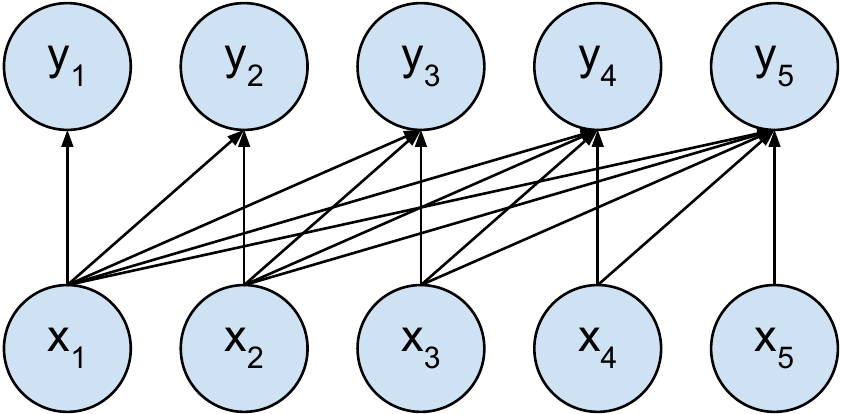}
           \label{fig:con:causal}
        }
        \hspace{1cm}
        \subfloat[lookback and lookahead of k=1 frames]{
        \includegraphics[width=0.40\linewidth]{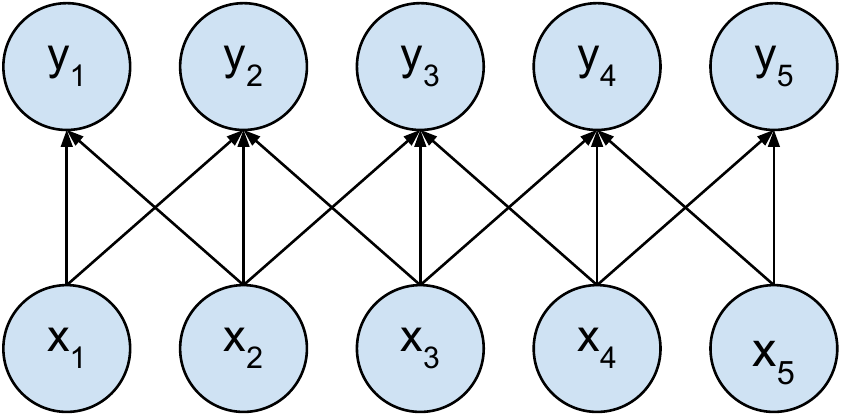}
           \label{fig:con:look}
        }
        \caption{Typical connectivity patterns at the sequence level for representation learning}
        \label{fig:tmp5}
    \end{figure}

The Transformer~\cite{transformers} is a good seq2seq candidate for this task and we choose it as a foundation for our system Taris. Unlike the recurrent neural network that uses causal connections between timesteps (Figure~\ref{fig:con:rnn}), the Transformer
allows feature contextualisation at the sequence level through self-attention, illustrated in Figure~\ref{fig:con:off}. This offline modelling strategy provides a theoretical upper limit of the segmentation performance. Furthermore, the self-attention connections in the Transformer block can be adjusted to allow causal modelling (Figure~\ref{fig:con:causal}) or non-causal modelling with a window (Figure~\ref{fig:con:look}). The window length is directly linked to the algorithmic latency of Taris and its accuracy, and we investigate this trade-off in Section~\ref{exp:counting}.

\section{Model architecture}
\label{sec:method}

\subsection{Encoding}

Taris takes as input a variable length sequence of audio vectors $\mathbf{a} = \{a_1, a_2, \ldots , a_{\mathbf{N}}\}$ and applies the Encoder stack of the Transformer model defined in~\cite{transformers}. Because of latency considerations, instead of the original full connectivity in Figure~\ref{fig:con:off}, we use the type displayed in Figure~\ref{fig:con:look}, with controlled look-back $e_{LB}$ and look-ahead $e_{LA}$ frames. We denote the outputs of the encoder as:
\begin{align}
    \mathbf{o_{A}}  = \mathsf{Encode}(\mathbf{a}, e_{LB}, e_{LA})
    \label{eq:a}
\end{align}

Next, we apply a sigmoidal gating unit on each encoder output $o_{A_i}$ to obtain a scalar score for each frame:
\begin{align}
    \alpha_i & = \mathsf{sigmoid}(o_{A_i} W_G  + b_G) \label{eq:alphai} \\
    & \nonumber \mathsf{where}\ \mathsf{sigmoid}(x) = \frac{1}{1 + \exp(-x)}, W_{G} \in \mathbb{R}^{\mathrm{h\,x\,1}}, b_{G} \in \mathbb{R}^{\mathrm{1}}
\end{align}

We assign to every single input frame $i$ a segment index $\hat{w_i}$ by taking the \emph{cumulative sum} of $\alpha$ and applying the \textit{floor} function on the output:
\begin{align}
    \hat{w_i} = \left\lfloor\sum_{j=1}^{i} \alpha_j  \right\rfloor
\end{align}

Namely, the first predicted segment is delimited by a cumulative sum of $\alpha$ between 0 and 1, the second segment by the same quantity between 1 and 2, and so on.

\subsection{Decoding}

During training, the Decoder stack receives the labelled grapheme sequence $\mathbf{y} = \{y_1, y_2, \ldots, y_{\mathbf{L}}\}$, made of English letters and the unique word delimiter SPACE. We assign to every grapheme $k$ a word index $w_k$ by leveraging the SPACE tokens in the labelled sequence:
\begin{align}
    w_k = \sum_{j=1}^{k} (y_j == \mathsf{SPACE})
\end{align}

Thus, whereas symbolic segmentation of speech uses a unique SPACE token to separate words, acoustic segmentation flags word boundaries by tracking the frame locations where the partial sum of the word counting signal $\alpha_i$ passes to the next integer value.

We modify the decoder-encoder connectivity of the Attention layer of~\cite{transformers} to allow our decoder to perform soft-alignment over a \emph{dynamic} window of segments estimated by the encoder. More precisely, we only allow those connections for which the following condition is met:
\begin{align}
    \mathbf{V} =  \widehat{W}_{ik} \leq (W_{ik} + d_{LA})\ \mathbf{and}\ \widehat{W}_{ik} \geq (W_{ik} - d_{LB})
    \label{eq:valid}
\end{align}

In~\eqref{eq:valid}, $d_{LA}$ and $d_{LB}$ denote the number of segments the decoder is allowed to look-ahead and look-back respectively. The $W$ and $\widehat{W}$ matrices are obtained from the $w$ and $\hat{w}$ arrays by applying the tile operation, which repeats one sequence for a number of times equal to the length of the other one. In more detail, $\mathbf{V}$ is a 2D matrix $\in \mathbb{R}^{\mathrm{N\,x\,L}}$ that defines the admissible connections between any decoder timestep and any encoder timestep, acting as a bias on the decoder-encoder attention. Setting $\mathbf{V}$ as a matrix of ones recovers the original Transformer model. The extension to 3D tensors that include the batch dimension is straightforward, offering Taris efficient minibatch training and inference.

The decoder implements a traditional character level auto-regressive language model that predicts the next grapheme in the sequence conditioned on all the previous characters and the dynamic audio context vector $c_k$:
\begin{align}
    c_k & = \mathsf{Attention}(\mathtt{keys=}\mathsf{o_{A}},\mathtt{query=} o_{D_{k-1}}, \mathtt{mask=}V) \\
    o_{D_k} & = \mathsf{Decode}(y, c_k) \\
    p_k & \equiv P(y_{k} | c_k, y_{1 : k-1}) = \softmax(o_{D_k} W_{v} + b_v) \label{eq:vocab} \\
    \nonumber & \mathsf{where} \ W_{v} \in \mathbb{R}^{\mathrm{h\,x\,v}}, b_{v} \in \mathbb{R}^{\mathrm{v}} 
\end{align}

In~\eqref{eq:vocab}, $v$ is the vocabulary size of 28 tokens.
We measure the difference between the estimated word sum $\Sigma \hat{w} = \sum_i \alpha_i$ and the true word count $|w| = \sum_k (y_k == \mathsf{SPACE})$ as:
\begin{align}
    Word\ Loss & = (|w| - \Sigma \hat{w})^2 \label{eq:wloss}
\end{align}

We define the training loss as:
\begin{align}
     CE\ Loss & = \frac{1}{L} \sum_k -y_k \log (p_k) \\
    \mathsf{Loss} & = CE\ Loss + \lambda\ Word\ Loss
\end{align}

In our experiments we used a scale factor $\lambda = 0.01$ found empirically. The self attention connections of the auto-regressive Decoder are causal as depicted in Figure~\ref{fig:con:causal}.


Taris requires a negligible overhead in parameters (given by the $W_G$ and $b_G$ variables in \eqref{eq:alphai}) and operations (equations \eqref{eq:alphai}-\eqref{eq:valid}) over the original Transformer.
\section{Why learn to count words}
\label{sec:discussion}

Proper lexical segmentation of speech depends on context and semantics, as commonly illustrated by the example \emph{how to wreck a nice beach} sounding similar to \emph{how to recognise speech}. Thus, strategies incrementally scanning for hard boundaries~\cite{li2019acs,Hou2020, dong2019cif} are less suited to word units, prompting~\cite{dong2019cif} to perform beam search on the entire sequence of sub-word tokens estimated from each segment. Instead, Taris has to develop intrinsic word counting mechanisms. One plausible strategy is to incrementally gather lexical evidence at the sub-word level, and learn to represent boundary-informative acoustic cues on a manifold where they can be accumulated.

\begin{figure}
    \centering
    \includegraphics[width=1.0\linewidth]{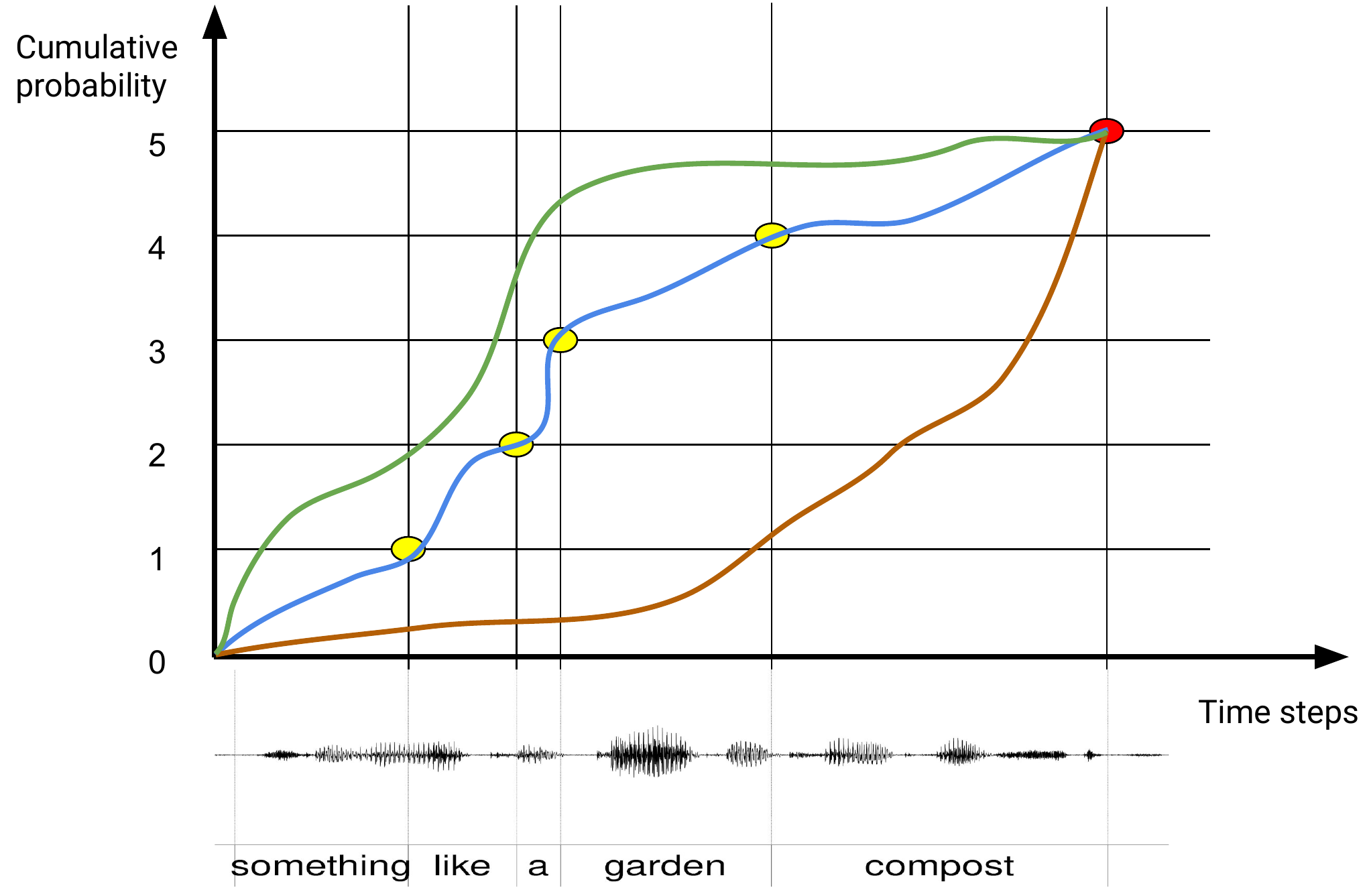}
    \caption{Illustration of the word counting problem. Given a spoken phrase \emph{something like a garden compost}, the network has to reach the correct word count 5 taking any path starting in the bottom left corner. The blue path allows an easy segmentation of speech into words. The green and dark orange paths are possible, but do not facilitate segmentation.}
    \label{fig:word_segment}
\end{figure}

We conjecture that learning the ability to count words facilitates the segmentation of speech into words, and we discuss below our intuition behind it. In Figure~\ref{fig:word_segment} we illustrate the word counting sub-problem to be solved by the network. Starting in the bottom left corner, the network predicts scores for every audio frame in the sentence, and the cumulative sum is promoted get as close as possible to the total word count, shown with a red circle. There is a very large number of paths that can be taken to reach the target count. However, when trained on large amounts of naturally distributed speech, we predict that Taris converges towards genuine word segmentation by having the cumulative sum cross all the intermediate word boundaries shown with yellow circles. In other words, the network may learn to self-normalise the accumulated probabilities for each word regardless of their length or cued structure.

We believe it suffices to train a system with the right amount of speech data, with the following intuition. As words appear in multiple contexts throughout a dataset, learning to count words may then have a normalisation effect on the fraction of $\Sigma \hat{w}$ allocated to each word in a sentence. Each word unit will approach a unitary mass allocation as its acoustic realisation is seen more often in multiple contexts. For the less frequent words, the correct allocation may happen by marginalisation if the sentences they appear in contain relatively more frequent words. Loosely speaking, it is the task of solving a system of linear equations where the variables are the partial sums corresponding to the acoustic frames between two consecutive estimated boundaries. 

Since we do not explicitly model the pauses between words, and the convergence towards the segmental behaviour is a mathematical conjecture without analytic proof for now, it is likely to observe deviations in practice on learnt solutions. However, Taris does not require a very strict approximation of word boundaries to function correctly. Instead, it is sufficient to just avoid frequent under- and over-segmentation, as it directly impacts the model's latency. 
\section{Experiments and Results}

We first conduct our experiments on the audio part of the unconstrained speech dataset LRS2~\cite{lrs2} for rapid prototyping, and on the 100h partition of LibriSpeech~\cite{panayotov2015librispeech} for empirical validation at a larger scale. To extract audio features $\mathbf{a}$ in Equation~\ref{eq:a}, we apply the log scale Short-time Fourier Transform on the waveform inputs, following same procedure as in~\cite{9035650} for noise corruption at 10, 0, and -5 db.

Our implementation of Taris is forked from the official Transformer model in TensorFlow 2~\cite{tf_transformer}. We train our LibriSpeech models for a total of 500 epochs at an initial learning rate of 0.001, decayed to 0.0001 after 400 epochs. The training time is approximately 200 seconds for a single epoch of LibriSpeech 100h on an Nvidia Titan XP GPU. The LRS2 models were trained with the same learning rates for 100 and 20 epochs respectively, on each noise level.

\subsection{Neural network details}
Our models use 6 layers in the Encoder and Decoder stacks, a model size $d_{model} \equiv h = 256$, a filter size $d_{ff} = 256$, one attention head, and 0.1 dropout on all attention weights and feedforward activations. The models occupy 25 MB on disk.

\subsection{The End-of-sentence (EOS) token}

During our initial experiments, we noticed that traditional evaluation and training strategies in neural speech recognition are commonly misusing the EOS token, making it difficult to evaluate online systems. The commonly used ASR datasets are a collection of variable-length utterances, and the system's accuracy is computed for each utterance using an edit distance based algorithm. These utterances are often fragments from full spoken sentences, such as the one illustrated in Figure~\ref{fig:word_segment}, that were cropped using voice activity detection algorithms (e.g. in LRS2), and sometimes the fragmentation includes the ending and the start of two consecutive sentences, with the punctuation removed from the ground truth transcription (e.g. in LibriSpeech). In other words, the ASR system does not receive full sentence units, and cannot develop the linguistic notion of an end of sentence. In our experiments it became obvious that one way the ASR model differentiates between an EOS token and a word delimiter (SPACE) likely comes from the apriori knowledge of the sentence length, and that EOS becomes more likely as the decoder-encoder alignment distribution advances towards the last remaining audio frames in the sentence. The aspect above becomes problematic in an online setting, as the decoder is fed with a limited acoustic context. Given the nature of the dataset utterances, an online decoder does not know when to stop the decoding process, as EOS cannot be estimated even spuriously anymore. Online decoding would often stop after just a few words in an utterance, biasing the accuracy on longer sentences.

To circumvent this problem, we made two important changes to the traditional model. First, we replaced the EOS token in the labels, which cannot be predicted reliably, with the SPACE token. Second, we modified the stopping condition of the beam search inference decoder as follows: instead of stopping when all beams reach the EOS token, it now stops when the decoder predicts as many words as there were estimated by the audio encoder. This new strategy is mostly beneficial to the evaluation procedure, but should also be useful in practice as it allows the decoder to emit a controllable number of words. With this change, we are able to evaluate the error rate of Taris on full test sentences for which we lack any label alignments.

\subsection{Learning to count words}
\label{exp:counting}
\begin{figure}[t]
        \centering
        \subfloat[Word Loss on LRS2]{
           \includegraphics[width=0.9\linewidth]{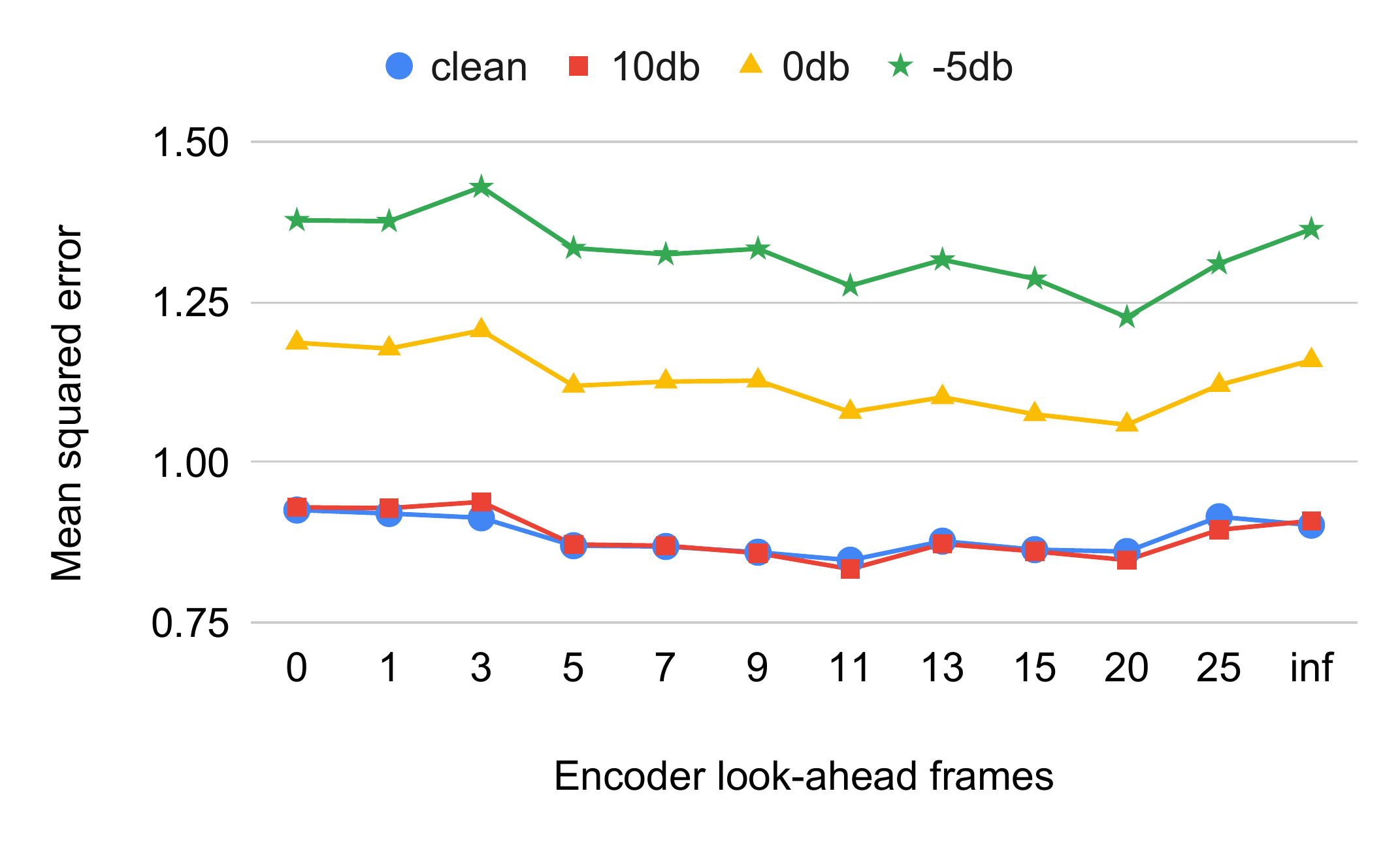}
           \label{fig:wc_loss}
        }
        
        \subfloat[Character Error Rate on LRS2]{
        \includegraphics[width=0.9\linewidth]{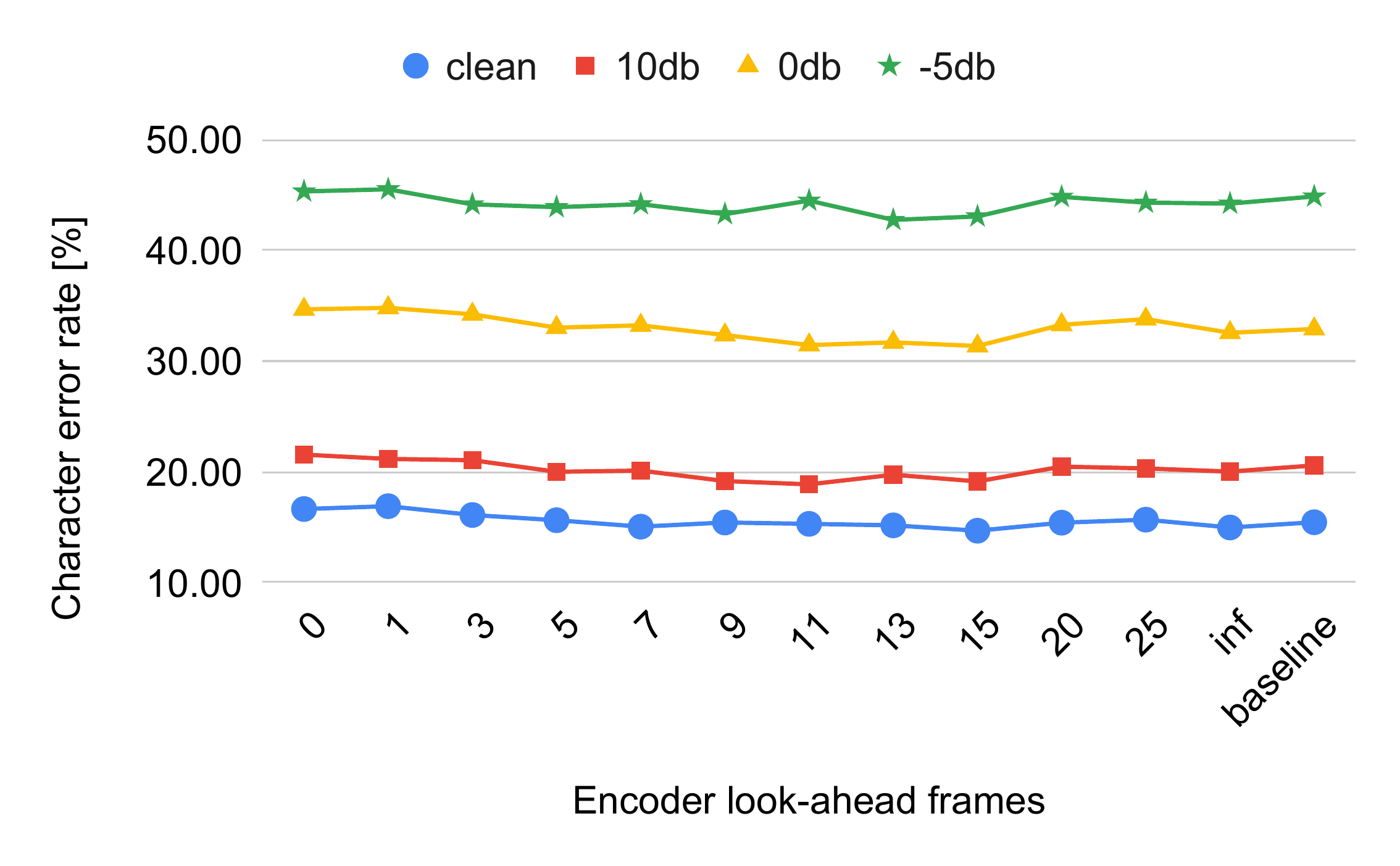}
           \label{fig:cer_offline}
        }
 
        \caption{Offline system evaluation for an increasing length of feature contextualisation in the encoder}
        \label{fig:wcperf}
    \end{figure}

We first investigate to what extent a sequence to sequence Transformer model can learn to count the number of words from audio data on LRS2. We train multiple models and gradually increase the number of encoder look-ahead frames $e_{LA}$ to measure the variation of the $Word\ Loss$ as more future context becomes available. We see in Figure~\ref{fig:wc_loss} that the mean squared word count error is sub-unitary in clean speech and 10db noise, i.e. the estimated count is less than one word away from truth. This suggests that words can be counted relatively well from acoustic speech. In addition, using a future context length of 11 frames offers the lowest counting error under all noise conditions. In Figure~\ref{fig:cer_offline} we plot the mean Character Error Rate achieved by all our systems, including the offline Transformer baseline without the auxiliary Word Loss, and we observe no significant difference, with the 95\% confidence intervals of the mean errors between 1\% and 1.4\%. Therefore, this auxiliary task is not detrimental to the original accuracy obtained on LRS2 using only the cross-entropy loss.

 \subsection{Online ASR decoding}
 \label{exp:online_decoding}
 
The decoder in our previous experiment had access to the entire encoder memory. For our online model we opt for an encoder lookahead $e_{LA}$ of 11 frames and infinite lookback $e_{LB} = \infty$, as we showed in Section~\ref{exp:counting} that there are diminishing gains beyond this threshold. This roughly corresponds to an encoding latency of 330 msec for each encoder layer.
 
\begin{figure}[t]
    \centering
    \includegraphics[width=\linewidth]{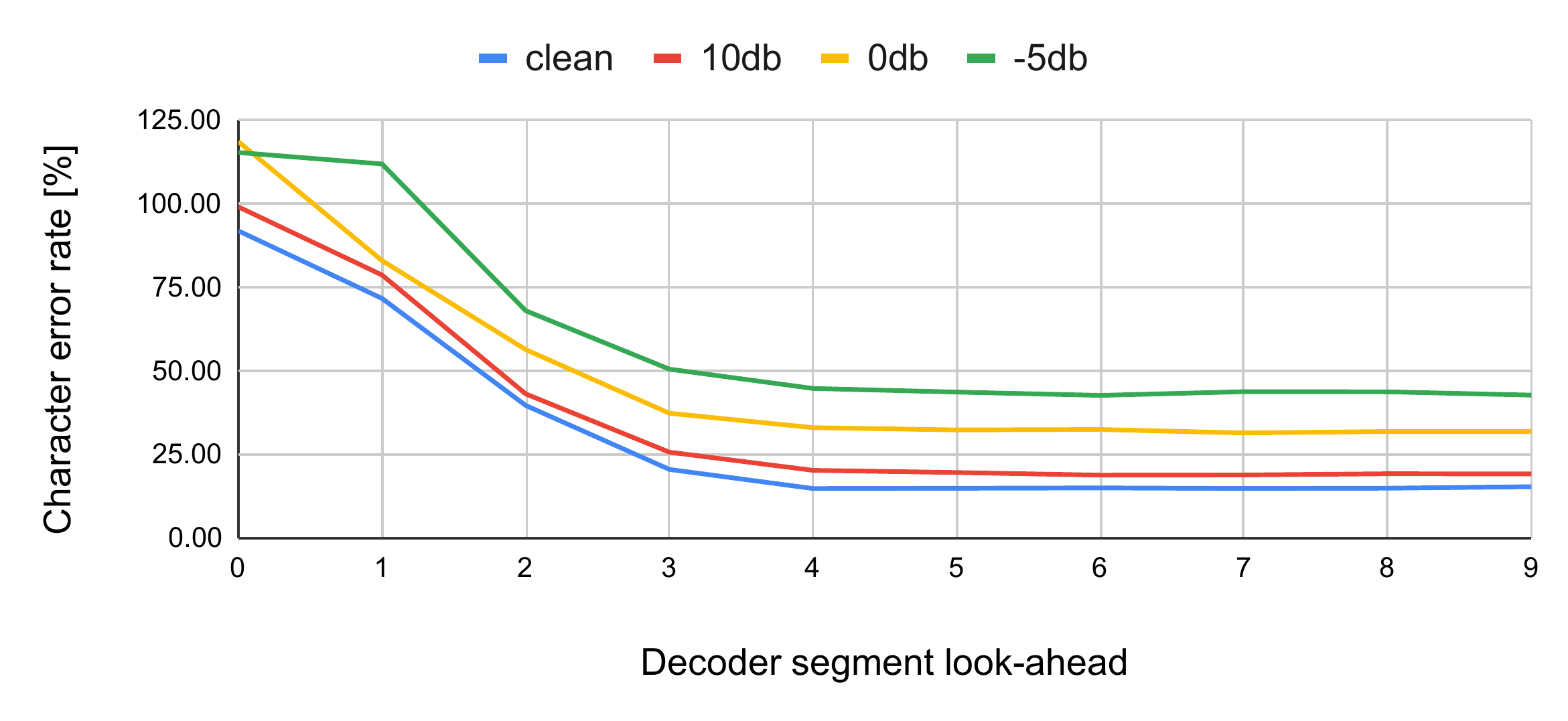}
    \caption{Online decoding error rate on LRS2. We fix $e_{LB} = \infty$, $e_{LA} = 11$ frames, $d_{LB} = \infty$ and we only allow $d_{LA}$ to vary.}
    \label{fig:ondec}
\end{figure}

    In this experiment we evaluate the error rate of Taris on LRS2 for an increasing number of decoder look-ahead segments $d_{LA}$, while setting the look-back value $d_{LB} = \infty$. For a practical online model it may be a good trade-off to limit the decoder look-back context to a single sentence when transcribing continuously. We plot the Character Error Rate in Figure~\ref{fig:ondec} for an increasing number of acoustic segments that the decoder is allowed to attend to. 
We notice that there are diminishing returns after a context look-ahead $d_{LA}$ of 4 words. The overall accuracy beyond this threshold is comparable to the offline systems shown in Figure~\ref{fig:cer_offline}.

\subsection{Evaluation on longer sentences}
\label{exp:librispeech}

In the previous experiments we have used the LRS2 dataset for rapid prototyping. However, since it contains many short sentences,
the potentially higher decoding error rate of Taris on the longer sentences might have little effect on the reported average error rate. We re-train and evaluate our models on the 100 hour clean partition of the LibriSpeech dataset, displaying the mean error and 95\% confidence interval (CI) around the mean in Table~\ref{tab:avt}.

First, we notice that the systems achieve an error rate similar to the one obtained on LRS2, despite the increased amount of data, suggesting that further gains are possible for larger model sizes. We also notice that the word loss can be slightly detrimental to the overall accuracy for the same network capacity, particularly for the models with unbounded attention span. This prompts a deeper investigation into the interplay between the cross entropy and word counting losses, as our constant scale factor $\lambda$ is likely a less than optimal solution to this multitask problem.    

\begin{table*}[t]
  \caption{System evaluation on LibriSpeech 100h clean partition.}
  \label{tab:avt}
  \centering
  \begin{tabular}{l|l|l|l|l|c c c}
    \toprule
    
    & \multicolumn{4}{c}{parameters} & \multicolumn{2}{c}{\multirow{1}{*}{\textbf{CER}}} & \multicolumn{1}{c}{\multirow{2}{*}{\textbf{Word Loss}}} \\
\textbf{Model} & $e_{LB}$   & $e_{LA}$  & $d_{LB}$  & $d_{LA}$  & mean [\%] & 95\% CI [\%]  & \multicolumn{1}{c}{} \\
\midrule
Transformer~\cite{transformers} & $\infty$   & $\infty$   & $\infty$  & $\infty$ & $13.37$ & 0.444 & $N/A$\\
Transformer + Word Loss & $\infty$   & $\infty$   & $\infty$  & $\infty$    & 14.64 & 0.451 & 0.92 \\
Taris : infinite look-back & $\infty$    & 11   & $\infty$    & 5    & 15.70 & 0.451 & 1.12 \\
Taris : finite look-back & 11 & 11 & 5 & 5 & 13.83 & 0.451 & 0.76\\
    \bottomrule
  \end{tabular}
  
\end{table*}

\begin{figure}[h]
        \centering
        \subfloat[LRS2]{
           \includegraphics[width=0.9\linewidth]{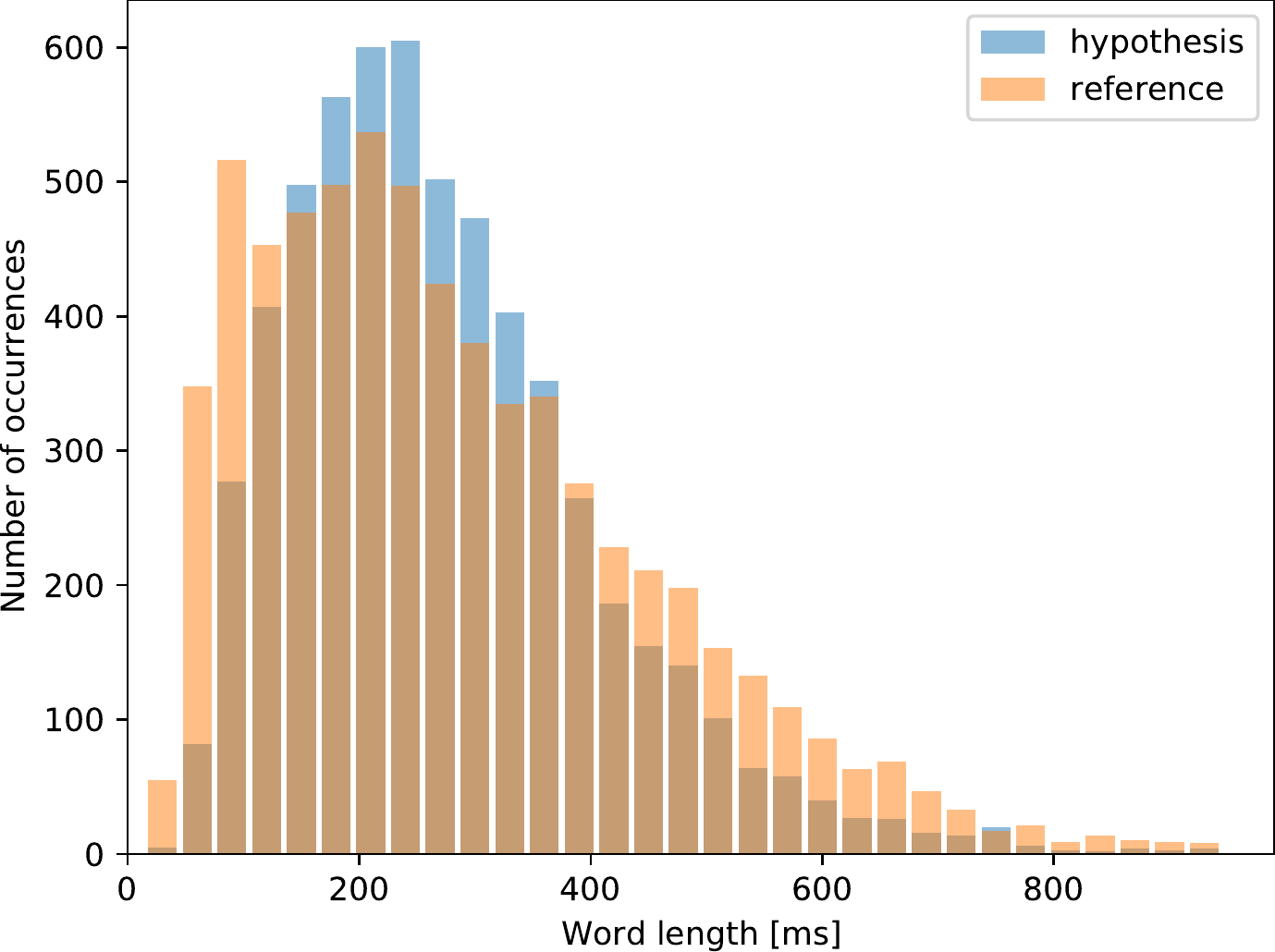}
           \label{fig:hist_lrs2}
        }
        
        \subfloat[LibriSpeech]{
        \includegraphics[width=0.9\linewidth]{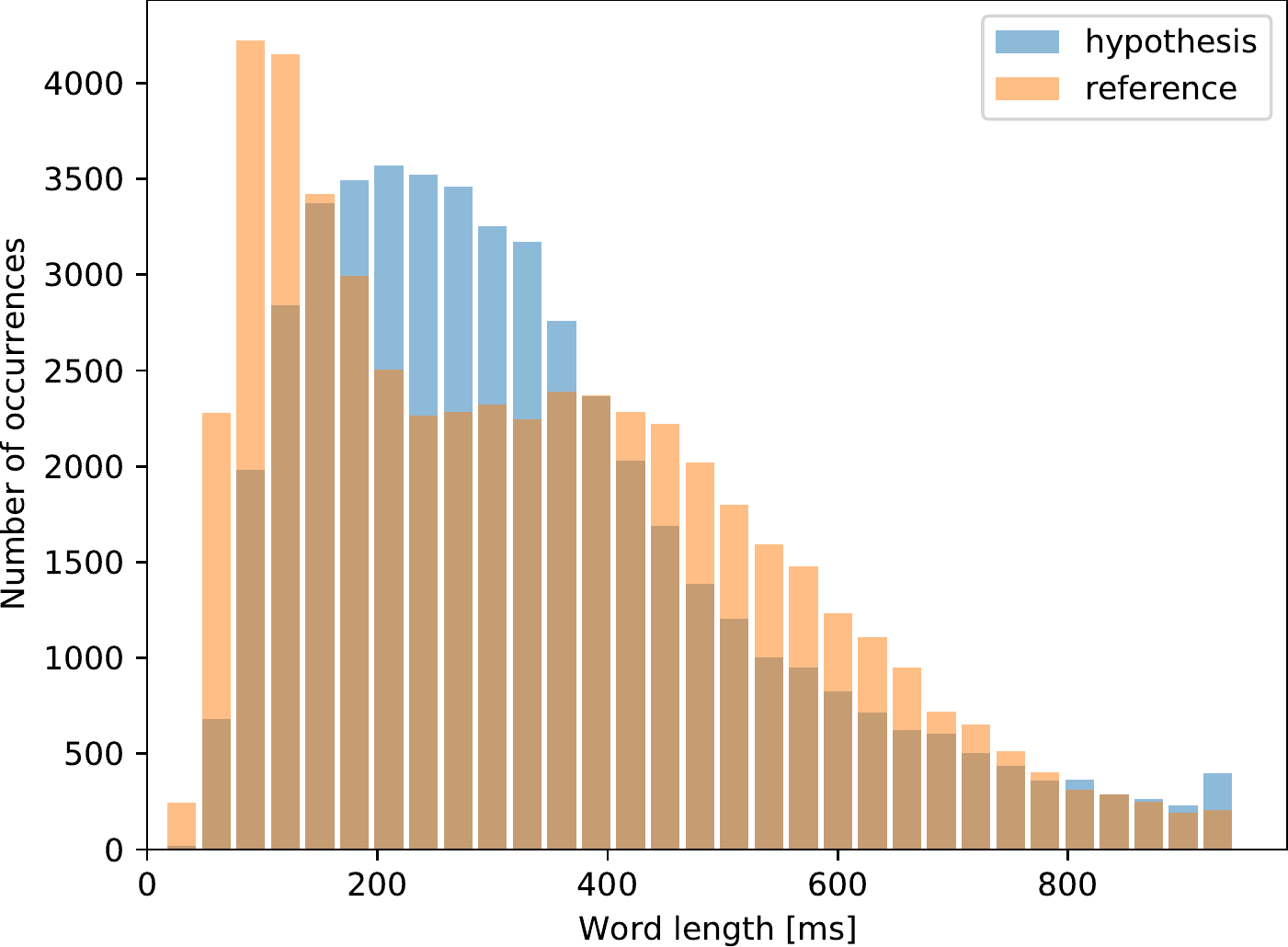}
           \label{fig:hist_libri}
        }
        \caption{Segmentation length distribution (in milliseconds) of Taris compared to the reference provided by the Montreal forced aligner}
        \label{fig:histograms}
    \end{figure}

Next, we compare the distributions of the segment lengths estimated by Taris and those estimated with the pre-trained Montreal forced aligner~\cite{McAuliffe2017}, both plotted in Figure~\ref{fig:histograms}. Not only are the histograms highly overlapped, but the one produced by Taris is in line with the average speaking rate of read speech. The small differences between the reference and hypothesis are likely owed to the short silences between words which were excluded from the reference, whereas Taris does not explicitly model silences and includes them into segments. Latency has not received sufficient consideration in prior work to facilitate a direct comparison, as systems were trained with offline encoders~\cite{moritz2019} or large receptive fields~\cite{Hou2020}, relied on beam search over the output distribution~\cite{dong2019cif}, or used phoneme units~\cite{jaitly2016}. Very recent work combining a weaker online model with an offline rescorer~\cite{sainath2020} that allows to revise online hypotheses with a final hypothesis introduces the notion of end-pointing latency. The offline rescorer is triggered after the utterance has been determined to be finished, that is when a threshold period has elapsed after a suspected end of utterance without further speech activity. Since our model is fully online and does not have to wait to rescore, this metric is not applicable to our system.

    \subsection{Evaluation on Mandarin speech}
    
    Since the word segmentation strategy in Taris is tailored for English, we are interested in extending the principle to Mandarin speech. Unlike English, Mandarin is characterised by a low number of morphemes per word and has almost no inflectional affixes, being considered a highly analytic language. In addition, the commonly used writing system belongs to the \emph{scriptio continua} style, with no delimitation between words. On these grounds, we make a structural change in Taris: instead of learning to count words (spaces between them), we let the system count the number of characters, similar to the quantity loss used in~\cite{dong2019cif}. This would drive the system to segment the acoustics associated with each character, which is almost always equivalent to a syllable.
    
    For our experiment we use the Aishell-1 dataset~\cite{Bu2017}, which contains 165 hours of fluent speech recordings from 400 speakers coming mostly from the Northern area of China, and covers a broad range of topics. The transcription file comprises an inventory of 4333 characters, which will determine the final output size of the decoder. Since the labels also include candidate blank spaces between words, we also evaluate Taris at the word level as we did on English. Despite the larger dataset size, we maintain the same Transformer size as before for faster prototyping. We label the different parametrisations of Taris as follows: WIDE: $e_{LB}=e_{LA}=11$ frames, $d_{LB}=d_{LA}=5$ segments,
    MEDIUM: $e_{LB}=e_{LA}=3$ frames, $d_{LB}=d_{LA}=5$ segments,
    NARROW: $e_{LB}=e_{LA}=2$ frames, $d_{LB}=d_{LA}=2$ segments.
    
    \begin{figure}[h]
        \centering
        \subfloat[Offline and online decoding error rate]{
           \includegraphics[width=0.9\linewidth]{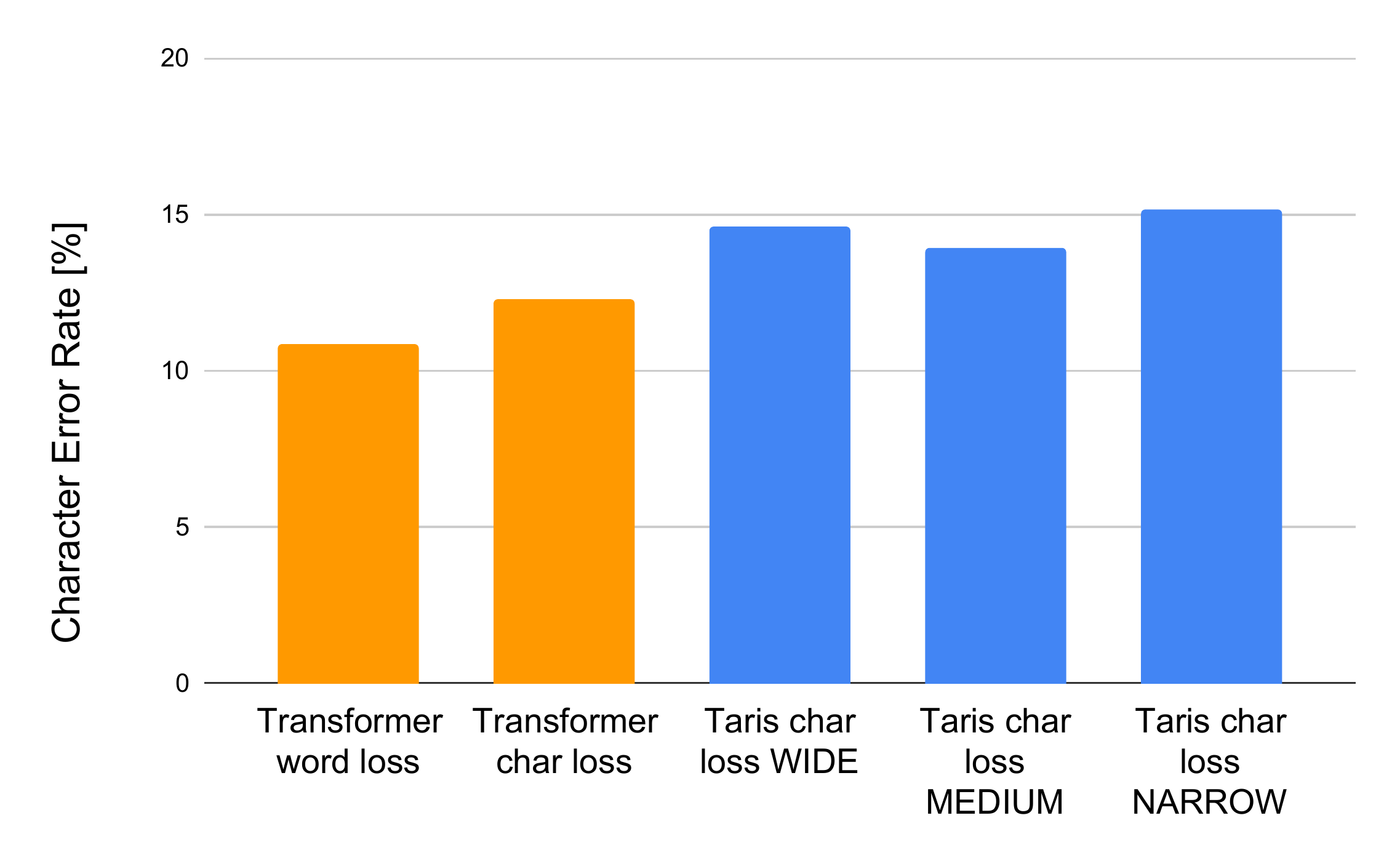}
            \label{fig:aishell_cer}
        }
        
        \subfloat[Unit counting loss]{
           \includegraphics[width=0.9\linewidth]{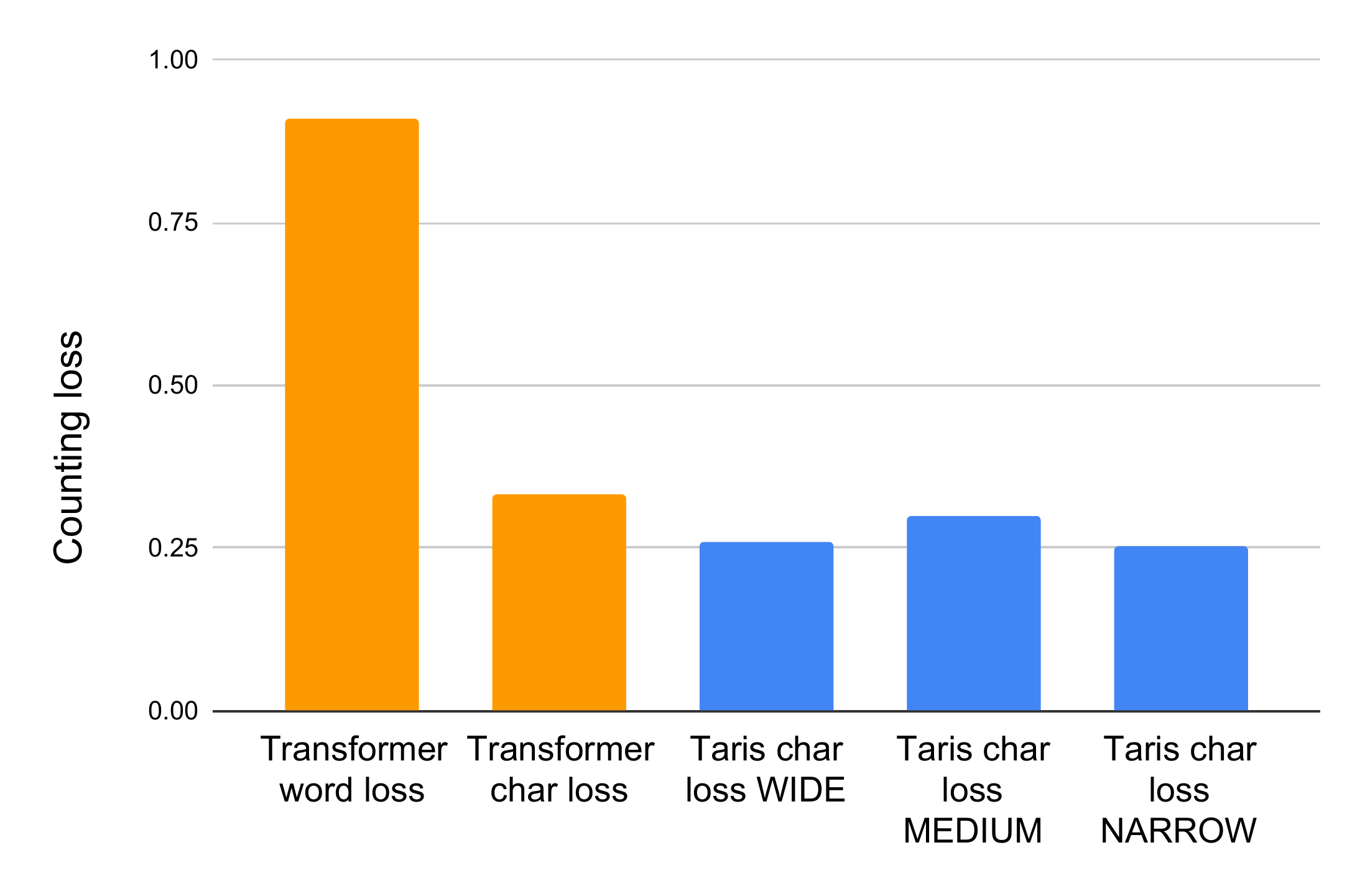}
            \label{fig:aishell_cl}
        }
    \caption{System Evaluation on Aishell-1}
    \label{fig:aishell}
    \end{figure}
    
    Figure ~\ref{fig:aishell} shows the error rates of the offline and online systems on the Aishell-1 corpus. Despite the small model size, the absolute decoding accuracy is comparable with the baseline results in~\cite{Bu2017} obtained with the Kaldi toolkit. We notice that Taris does a much better job at learning to count the number of characters than the number of words, with our NARROW model obtaining a counting error of 0.2538. Since a Chinese character almost always corresponds to a single syllable, this result suggests that syllables may be easier to segment in fluent speech than words. Furthermore, the syllable level segmentation allows both the encoder and the decoder of Taris to use relatively low context lengths and further reduce the overall latency. Not shown in the figure, the error rate of a \emph{word} counting Taris model is approximately 40\%, implying that a good unit segmentation is essential for online decoding.

\section{Conclusion}

We have proposed a simple, efficient, and fully differentiable solution for online speech recognition that does not require additional labels. Taris is inspired from early language acquisition in infants, and aims to segment a speech stream by learning to count the number of words therein. We show that our method matches the accuracy of an offline system once it listens to 5 dynamic segments. Lowering this latency remains a topic for exploration, e.g. by gradually reducing the look-ahead parameter $d_{LA}$ later in training, explicitly modelling silences, or investigating the role of context, grammar, and semantics in lexical recognition.


Generalising to sentences of different lengths from the ones seen in training has recently been identified as a major problem for neural online speech recognition systems~\cite{9003854, 9003913}. By modelling only the local relationships in speech through finite look-back and look-ahead, we preserve the same property of the Neural Transducer~\cite{jaitly2016} to effectively decouple the sentence length from the learnt representations, while allowing adaptive segments and simpler training.

It can be argued that Taris exploits human knowledge of the speech signal structure and embeds the concept of words and the local acoustic relationships, instead of being a more generic, self-organising neural network. Yet, the local processing of speech is merely the one dimensional equivalent of local convolutions applied to images, where the objects are replaced by words. Moreover, one-dimensional convolutions are commonly used in speech recognition~\cite{hamid2014, pratap2019, jasper, kriman2020}. Given their major impact in research despite their lack of invariance to orientation, scaling, or even small perturbations, there is still much to be learned from engineered models in the pursuit of artificial general intelligence.

\section{Acknowledgement}

This research was conducted with the financial support of Science Foundation Ireland under Grant Agreement No. 13/RC/2106 at the ADAPT SFI Research Centre at Trinity College Dublin, The University of Dublin, Ireland. The ADAPT SFI Centre for Digital Media Technology is funded by Science Foundation Ireland through the SFI Research Centres Programme and is co-funded under the European Regional Development Fund (ERDF) through Grant \# 13/RC/2106. Our work is supported by a Titan Xp GPU grant from NVIDIA.

\newpage

\bibliographystyle{IEEEbib}
\bibliography{mybib}

\end{document}